\documentclass[useAMS,usenatbib]{mn2e}

\usepackage{graphicx}
\usepackage{subfig}
\usepackage{multirow}
\usepackage{array}

\newcommand{\xmm}{{\em XMM-Newton}}
\newcommand{\chan}{{\em Chandra}}

\title[GTC observations of $\gamma$-ray pulsars]{Observations of one young and three middle-aged $\gamma$-ray pulsars with the Gran Telescopio Canarias}
\author[R. P. Mignani, V. Testa, N. Rea,   et al. ]
{\parbox{\textwidth}{R. P. Mignani$^{1,2}$\thanks{E-mail: mignani@iasf-milano.inaf.it}, 
V. Testa$^{3}$,
N. Rea$^{4,5,6}$, 
M. Marelli$^{1}$,
D. Salvetti$^{1}$,
D. F. Torres$^{4,6,7}$,
E. De O\~{n}a Wilhelmi$^{4,6}$
} 
\\ \\
$^{1}$ INAF - Istituto di Astrofisica Spaziale e Fisica Cosmica Milano, via E. Bassini 15, 20133, Milano, Italy\\
$^{2}$ Janusz Gil Institute of Astronomy, University of Zielona G\'ora, ul Szafrana 2, 65-265, Zielona G\'ora, Poland \\
$^{3}$ INAF - Osservatorio Astronomico di Roma, via Frascati 33, 00040, Monteporzio, Italy \\
$^{4}$ Institute of Space Sciences (ICE, CSIC), Campus UAB, Carrer de Magrans s/n, 08193 Barcelona, Spain \\
$^{5}$ Anton Pannekoek Institute for Astronomy, University of Amsterdam, Postbus 94249, NL-1090 GE Amsterdam, the Netherlands \\
$^{6}$ Institut d?Estudis Espacials de Catalunya (IEEC), 08034 Barcelona, Spain \\
$^{7}$ Instituci\'o Catalana de Recerca i Estudis Avan\c{c}ats(ICREA), E-08010 Barcelona, Spain
 }

\begin{document}

\date{Accepted 1988 December 15. Received 1988 December 14; in original form 1988 October 11}

\pagerange{\pageref{firstpage}--\pageref{lastpage}} \pubyear{2002}

\maketitle

\label{firstpage}

\begin{abstract}
We used the 10.4m Gran Telescopio Canarias to search for the optical counterparts to four isolated $\gamma$-ray pulsars, all detected in the X-rays by either \xmm\ or \chan\ but not yet in the optical.  Three of them are middle-aged pulsars -- PSR\, J1846+0919 (0.36 Myr), PSR\, J2055+2539 (1.2 Myr), PSR\, J2043+2740 (1.2 Myr) -- and one, PSR\, J1907+0602, is a young pulsar (19.5 kyr).
For both PSR\, J1907+0602  and PSR\, J2055+2539 we found one object close to the pulsar position. However, in both cases such an object cannot be a viable candidate counterpart to the pulsar. For PSR\, J1907+0602, because it would imply an anomalously red spectrum for the pulsar and for PSR\, J2055+2539  because the pulsar would be unrealistically bright ($r'=20.34\pm0.04$)  for the assumed distance and interstellar extinction. For PSR\, J1846+0919, we found no object 
sufficiently close to the expected position to claim a possible association, whereas for PSR\, J2043+2740 we confirm our previous findings that the object 
nearest to the pulsar position 
is an unrelated field star. 
We used our brightness limits ($g' \approx 27$), the first obtained with a large-aperture telescope  for both  PSR\, J1846+0919 and PSR\, J2055+2539, to constrain the optical emission properties of these pulsars and investigate the presence of spectral 
turnovers at low energies in their multi-wavelength spectra.
\end{abstract}

\begin{keywords}
stars: neutron -- pulsars: individual: 
\end{keywords}

\section{Introduction}

Isolated neutron stars (INSs) are intrinsically very faint in the optical domain, where 
they are detected only through the emission of synchrotron radiation from relativistic particles in their magnetosphere or through the 
emission of thermal radiation from their hot (temperatures $T\sim 10^{5}$--$10^{6}$ K)  surface (e.g., Mignani 2011). Both processes are seen to co-exist in middle-aged objects (0.1--1 Myr), which makes them  intriguing targets to simultaneously study the emission properties of 
both the neutron star  surface and magnetosphere.  
The copious detection of $\gamma$-ray pulsars by the {\em Fermi} Large Area Telescope (LAT; Atwood et al.\ 2009), 
with over 200 of them\footnote{{\texttt https://confluence.slac.stanford.edu/display/GLAMCOG/\\Public+List+of+LAT-Detected+Gamma-Ray+Pulsars}}  identified since the launch of
the satellite in 2008, makes them the largest class of identified Galactic $\gamma$-ray sources.  Recent reviews on the {\em Fermi} contribution in pulsar $\gamma$-ray astronomy are 
found in Caraveo (2014) and Grenier \& Hardings (2015). The harvest of pulsar detections in $\gamma$-rays has fostered the interest on their multi-wavelength studies: in radio, 
to detect radio pulsations from pulsars discovered through blind periodicity searches in $\gamma$-rays (e.g., Saz Parkinson et al.\ 2010) and assess the ratio between radio-loud and radio-quiet $\gamma$-ray pulsars, and in the X-rays and in the optical, ultraviolet, infrared, to characterise the spectral energy distribution (SED) and study the interplay between different emission mechanisms.

  \begin{table*}
\begin{center}
\caption{Coordinates (with the uncertainties in parentheses) and reference epoch of the {\em Fermi} pulsars discussed in this work. The spin-down parameters and inferred quantities have been collected from the ATNF pulsar catalogue (Manchester et al.\ 2005). }
\label{psr}
\begin{tabular}{lllcccccc} \hline
Pulsar 				     &$\alpha_{J2000} $	& $\delta_{J2000}$&    Epoch & P$_{\rm s}$	&    $\dot{P_{\rm s}}$ 	& $\tau$ & B & $\dot{E}$ \\ 
                                          &    $^{(hms)}$   &      $^{(\circ ~'~")}$   &  MJD &   (s) &  (10$^{-14}$s s$^{-1}$)    &  ($10^{5}$ yr) & ($10^{12}$ G)  & ($10^{35}$ erg cm$^{-2}$ s$^{-1}$)  \\ \hline
J1846+0919$^{C}$                     & 18 46 25.8  ($0\fs027$) &    +09 19 49.8  ($0\farcs45$) &  56090 & 0.225 & 0.993 & 3.6 & 1.51 & 0.34 \\
J1907+0602$^{C}$                         & 19 07 54.76 ($0\fs05$)&   +06 02 14.6 ($0\farcs7$)& 55555  & 0.106 & 8.682  & 0.195 & 3.08 & 28 \\
J2043+2740                    & 20 43 43.5 ($0\fs1$)   &  +27 40 56 ($1\farcs0$)    &   49773  & 0.096 & 0.127 &   12 & 0.35 &  0.56 \\ 
J2055+2539$^{C}$         & 20 55 48.96  ($0\fs05$) &  +25 39 58.78 ($0\farcs9$)     & 57277   & 0.319 & 0.408  & 12.4 & 1.16 & 0.049 \\ \hline  
\end{tabular}
\end{center}
$^{C}$ For the radio-quiet pulsars PSR\, J1846+0919 and PSR\, J2055+2539 and for PSR\, J1907+0602 the coordinates have been obtained with {\em Chandra} (Marelli et al.\ 2015; Mareli et al.\, in preparation; Kerr et al.\ 2015).
\end{table*}

Optical follow-ups of isolated $\gamma$-ray pulsars rely upon observational efforts with 8--10m class telescopes, and with the {\em Hubble Space Telescope} and, so far, 
yielded the identification of likely candidate counterparts to PSR\, J0205+6449 (Moran et al.\ 2013), PSR\, J1741$-$2054 (Mignani et al.\ 2016a), and PSR\, J2124$-$3358 (Rangelov et al.\ 2017).  
In Mignani et al.\ (2016b) -- hereafter Paper I -- we reported on the observations of three  {\em Fermi} pulsars in the northern hemisphere, carried out with the 10.4 m Gran 
Telescopio Canarias (GTC) at the  La Palma Observatory (Roque de Los Muchachos, Canary Islands, Spain). Here we report on the continuation of our program for 
targets observable in the summer semester.   Like in Paper I, we selected our targets according to the following criteria: visibility constraints, X-ray detection with either \chan\ or \xmm, energetic,  non extreme distance and extinction, and accurate reference position ($\la 1\arcsec$).
In particular, we observed the middle-aged ($\approx$0.3--1.2 Myr) $\gamma$-ray pulsars PSR\, J1846+0919,  PSR\, J2043+2740, and PSR\, J2055+2539. 
Identifying the optical counterparts of pulsars in this age range, with only three of them firmly identified so far\footnote{These are Geminga (Bignami et al.\ 1987), PSR\, B0656+14 
(Caraveo et al.\ 1994), and PSR\, B1055$-$52 (Mignani et al.\ 1997), with a candidate counterpart found for PSR\, J1741$-$2054 (Mignani et al.\ 2016a).}, is important to study the evolution of the pulsar multi-wavelength emission properties. In addition,  we observed the younger PSR\, J1907+0602 ($\approx 19.5$ kyr), which is very similar to the Vela pulsar in its spin-down parameters. The characteristics of these pulsars, spin period  P$_{\rm s}$,  its derivative ($\dot{P_{\rm s}}$), characteristic age ($\tau$), surface magnetic field ($B$), and spin-down energy ($\dot{E}$), are summarised in Table 1.\footnote{Characteristic age and surface magnetic field are defined as in Manchester et al.\ (2005).} 
In the course of our program, we also observed the mode-changing $\gamma$-ray pulsar PSR\, J2021+4026 in the $\gamma$ Cygni supernova remnant (Allafort et al.\ 2013) as a part of a multi-wavelength campaign and the results are reported in a separate publication (Razzano et al.\ in preparation).

PSR\, J1846+0919 (P$_{\rm s}$=0.225 s) is a radio-quiet pulsar, identified in $\gamma$-rays by Saz Parkinson et al.\ (2010). An X-ray counterpart to PSR\, J1846+0919 has been identified with {\em Chandra} in a short 15 ks observation (Marelli et al.\ 2015),  although it was not possible to detect X-ray pulsations. The pulsar remained undetected in radio follow-up observations (Frail et al.\ 2016).  Like for other radio-quiet pulsars, its distance has not been measured yet. However, from the fit between  the $\gamma$-ray luminosity $L_{\gamma}$,  computed for pulsars with measured distance,  and the spin-down energy $\dot{E}$ one can extrapolate an expected value for the $\gamma$-ray luminosity and infer a "pseudo distance"  from the ratio with the observed $\gamma$-ray flux $F_{\gamma}$  (see, e.g. Equation 2 in Saz Parkinson et al.\ 2010). For PSR\, J1846+0919, this method gives a "pseudo-distance" D$_{\gamma}$ of $\approx$1.4 kpc (Marelli et al.\ 2015), although lower values cannot be ruled out. No follow-up optical observations of PSR\, J1846+0919 has ever been carried out till the present work. Serendipitous data from the UK Infrared Deep Sky Survey (Lawrence et al.\ 2007) give no compelling limits on the pulsar flux (Marelli et al.\ 2015).

PSR\,  J1907+0602 (P$_{\rm s}$=0.106 s) has the highest $\dot{E}$ in our sample. It  was discovered as a $\gamma$-ray pulsar during a blind search in unidentified {\em Fermi} sources (Abdo et al.\ 2009) and, soon after,  very faint radio pulsations at 1.5 GHz were  detected from Arecibo (Abdo et al.\ 2010; Ray et al.\ 2011).  The radio dispersion measure (DM=82.1$\pm$1.1 cm$^{-3}$ pc) puts PSR\,  J1907+0602 at a distance D$_{\rm NE}$=3.2$\pm$0.6 kpc (Abdo et al.\ 2010), from the NE2001 model of the Galactic free electron density  (Cordes \& Lazio 2002).  A slightly lower value (2.58 kpc) is obtained from the most recent model of Yao et al.\ (2017), now assumed as a reference in the ATNF pulsar catalogue\footnote{http://www.atnf.csiro.au/people/pulsar/psrcat/} (Manchester et al.\ 2005). Owing to the pulsar faintness, however, its distance has not been confirmed yet by radio parallax measurements. PSR\,  J1907+0602 is also embedded in a large ($\sim 40\arcmin$) pulsar wind nebula (PWN) detected at TeV energies  but not in the X-rays (see, e.g. Abeysekara et al.\ 2016 and references therein). 
The pulsar was detected both by \chan\  (Abdo et al.\ 2010; Marelli et al.\ 2011), and  \xmm\ (Abdo et al.\ 2013) but no X-ray pulsations have been detected yet.  The first deep optical observations were obtained with the Very Large Telescope (VLT) by Mignani et al.\ (2016c) but no candidate optical counterpart was detected down to $V\sim 26.9$. Our new data complement at longer wavelengths those of Mignani et al.\ (2016c).

PSR\, J2043+2740 (P$_{\rm s}$=0.096 s) is a radio loud  (Ray et al.\ 1996)  $\gamma$-ray pulsar  (Pellizzoni et al.\ 2009; Abdo et al.\ 2010; Noutsos et al.\ 2011). Its X-ray counterpart was detected  by \xmm\  but no X-ray pulsations were detected (Becker et al.\ 2004).  Its  radio dispersion measure (DM=21.0$\pm$0.1 pc cm$^{-3}$; Ray et al.\ 1996) gives a distance D$_{\rm NE}$=1.8$\pm$0.3 kpc (1.48 kpc according to the model of Yao et al.\ 2017) but a value as small as few hundreds pc cannot be excluded from the low hydrogen column density towards the pulsar  ($N_{\rm H} \la 3.6 \times 10^{20}$ cm$^{-2}$; Abdo et al.\ 2013) and the correlation between $N_{\rm H}$ and distance (He et al.\  2013).  
The first, deep, optical observations of the PSR\, J2043+2740 field have been recently obtained by Testa et al.\ (2018) with the Large Binocular Telescope (LBT) but the pulsar was not 
detected down to $V\sim 26.6$ also owing to non-optimal atmospheric conditions.

PSR\, J2055+2539 (P$_{\rm s}$=0.319 s) is radio-quiet, also detected as a $\gamma$-ray pulsar in a blind search (Saz Parkinson et al.\ 2010). The pulsar has not been detected in follow-up radio observations (Frail et al.\ 2016). In X-rays, it was detected by \xmm\  (Marelli et al.\ 2016), which also measured X-ray pulsations for the first time.  \xmm\ also detected two few arcmin-long X-ray emission tails,  apparently originating from the pulsar and separated by an angle of $\sim 160^{\circ}$.  The results of detailed multi-wavelength investigations of these tails, including optical broad and narrow-band imaging,  will be reported in a separate paper (Marelli et al.\, in preparation). Although PSR \, J2055+2539 has the lowest  $\dot{E}$ in our sample, it is expected to be quite close. According to the "pseudo-distance" method, the pulsar should be at D$_{\gamma} \approx$0.6 kpc but it could be at a distance as low as $\approx 0.4$ kpc (Marelli et al.\ 2016). Then, PSR\, J2055+2539 would be one of the eight middle-aged $\gamma$-ray pulsars closer than $\sim$ 0.5 kpc (see Abdo et al.\ 2013), of which already three have been detected in the optical and one has a candidate counterpart.
Despite of this, no deep optical observations of PSR\, J2055+2539 have been reported so far.

This manuscript is structured as follows: observations, data reduction and analyses are described in Sectn. 2, while the results are presented and discussed in Sectn. 3 and 4, respectively. Conclusions follow.

\section{Observations}

\subsection{Observation description}

We observed the four pulsar fields with the GTC in August 2015,  and June, July 2016 under   programmes GTC12-15A and  GTC27-16A (PI. N. Rea).  The observations were performed in service mode  with the OSIRIS camera (Optical System for Imaging and low Resolution Integrated Spectroscopy). The instrument is equipped with a two-chip E2V CCD detector with a combined field--of--view of $7\farcm8 \times 7\farcm8$, which, however, is substantially affected by vignetting in Chip 1.The pixel size of the CCD is 0\farcs25 ($2\times2$ binning). In total, we took a minimum of three sequences of five exposures in the $g'$  ($\lambda=4815$ \AA; $\Delta \lambda=1530$\AA), $r'$ ($\lambda=6410$ \AA; $\Delta \lambda=1760$\AA) and $i'$   ($\lambda=7705$ \AA; $\Delta \lambda=1510$\AA) filters (Fukugita et al.\ 1996) with exposure times of either 145 or 155 s,  to minimise the saturation of bright stars in the field and remove cosmic ray hits.   Exposures were dithered by 20\arcsec\ steps in right ascension and declination but always keeping the targets close to the nominal aim point in Chip 2. 
A summary of the GTC observations is given in Table \ref{gtc}. Observations were performed in dark time and under photometric or clear sky conditions, with seeing mostly below 1\farcs0 and the targets at airmass always close to 1. Short (0.5--3 s) exposures of standard star fields (Smith et al.\ 2002) were also acquired each night to allow for photometric calibration in the AB system (Oke 1974)
and zero point trending\footnote{{\texttt www.gtc.iac.es/instruments/osiris/media/zeropoints.html}}, together with twilight sky flat fields, as part of the OSIRIS service mode calibration plan (Cabrera-Lavers et al.\ 2014).

\begin{table}
\centering
\caption{Summary of the GTC optical observations of the three {\em Fermi} pulsars in Table \ref{psr}. Columns list the observing date, band, the total  integration time (T), the average airmass and seeing, and the sky conditions during the exposures.}
\label{gtc}
\begin{tabular}{lccccc} \hline
 Date      &  Band & T   & airmass & seeing & sky \\
                   yyyy-mm-dd   &    &  (s)  & & & \\ \hline
                              \multicolumn{6}{c}{J1846+0919} \\ \hline 
		               2016-06-06   &  $r'$ & 2175  & 1.07  & 1\farcs2  & clear\\ 
			                      &  $g'$ & 2175 & 1.11  & 0\farcs7  & phot \\ 
                                2016-06-07  &  $g'$ & 2175 & 1.08  &  1\farcs0 & phot \\ 
			                     &  $r'$ & 2175 &  1.15 & 0\farcs7  & phot \\ \hline
			         \multicolumn{6}{c}{J1907+0602} \\ \hline 
            2016-06-07   &  $r'$ & 2175 & 1.13 & 0\farcs6 & phot\\ 
                                                                 &  $i'$ & 2175 & 1.08 & 0\farcs7  & phot \\ 
			       2016-06-08 &  $r'$ & 2465 & 1.11 &  0\farcs6  & phot \\ 	             
			                  &  $i'$ & 2175 & 1.08 &  0\farcs6  & phot  \\ \hline
					  \multicolumn{6}{c}{J2043+2740} \\ \hline 
           2016-07-01  &  $g'$ & 2175 & 1.04  & 0\farcs9 & clear \\ 
                                           &  $r'$ & 2320 & 1.04 & 0\farcs7 & clear \\ 
 			       2016-07-03   &  $g'$ & 2175 & 1.08 & 0\farcs9 & clear \\ 
			                  &  $r'$ & 2175 & 1.18  & 1\farcs2 & clear \\ \hline  
			       \multicolumn{6}{c}{J2055+2539} \\ \hline 
	     	2016-07-03 & $g'$ &	2175 & 1.04 &1\farcs0 & clear  \\
                                                               & $g'$  & 2175 & 1.05 & 1\farcs1 & clear \\	
                               2015-08-24 & $r'$   & 2480 &  1.08 & 0\farcs7 & clear \\ \hline
\end{tabular}
\vspace{0.5cm}
\end{table}

 \subsection{Data reduction and calibration}

As done in Paper I, we reduced our data  (bias  subtraction, flat-field correction) using standard procedures in the {\sc IRAF}\footnote{IRAF is distributed by the National Optical 
Astronomy Observatories, which are operated by the Association of Universities for Research in Astronomy, Inc., under cooperative agreement with the National Science Foundation.} 
package {\sc ccdred}. We, then, aligned and co-added single dithered exposures  using the task {\tt drizzle} applying a cosmic-ray filter.

As done in, e.g. Testa et al.\ (2018), we analysed the images using the {\sc DAOPHOT-II} package (Stetson 1994). We applied its usual routines to run the object detection, fit the point spread function (PSF), and compute 
object photometry in an aperture of radius equal to the fitted image PSF. We adopted this technique because it has been found to give more robust results for faint objects. After these steps we computed the aperture correction on a subset of relatively bright isolated stars
and applied this factor to the instrumental magnitudes measured with  {\sc DAOPHOT-II}.
We then calibrated the aperture-corrected instrumental magnitudes by applying the photometric zero points
computed from the observations of the standard star fields mentioned in Sectn.\ 2.1. For the airmass correction we adopted the extinction coefficients measured 
for the La Palma Observatory\footnote{{\texttt  www.ing.iac.es/Astronomy/observing/manuals/ps/tech\_notes/tn031.pdf}}.
 We cross-checked our photometry calibration against the magnitudes of  stars selected from the Sloan Digital Sky Survey (SDSS; York et al.\ 2000) catalogues, 
which are in the Fukugita et al.\  (1996) filters, for the two pulsars for which there are stars in the SDSS catalogue, namely PSRJ2043+2740 and PSRJ2055+2539. 
We have done the comparison against $\sim$200 stars  in the PSR\, J2043+2740 field and $\sim$30
in the less crowded field of PSR\, J2055+2539, all in the magnitude range $19.5 < g', r' < 22$, and found agreement at better than 0.05 magnitudes in all filters.

We computed the astrometric solution on the GTC images using the {\em wcstools}\footnote{{\texttt http://tdc-www.harvard.edu/wcstools}} 
software package, matching the sky and pixel coordinates of stars selected from the Two Micron All Sky Survey (2MASS) All-Sky Catalog of Point 
Sources (Skrutskie et al.\ 2006).  After iterating the procedure applying a $\sigma$-clipping to filter out obvious mismatches, high-proper 
motion stars, and false detections, we obtained mean residuals of $\sim 0\farcs2$ in the radial direction, using at least 30 bright, but non-saturated,  
2MASS stars. The uncertainty on the centroids of the reference stars is negligible compared to the pixel scale of the OSIRIS images (0\farcs25). To the mean 
residual of the sky--to--pixel transformation we added in quadrature the uncertainty on the registration on the astrometry reference frame (Lattanzi et al.\ 1997), 
which is $\sigma_{tr}$=$\sqrt{n/N_{S}}\sigma_{\rm S} \la  0\farcs07$, where $n$=5 is the number of free parameters in the sky--to--image transformation 
(rotation angle, x, y offsets and scales), $N_{S}$ the number of stars used for the astrometry calibration, and $\sigma_{\rm S}$ is  their mean absolute position error, 
which is $\sim 0\farcs2$ for 2MASS stars of magnitude $15.5 \le K \le 13$ (Skrutskie et al.\ 2006). 
Thus, the overall accuracy on our absolute astrometry is $\sim$0\farcs2, which largely accounts for the uncertainty on the link of 2MASS to the 
International Celestial Reference Frame (0\farcs015; Skrutskie et al.\ 2006).

\section{Results}

We used the most recent pulsar positions (Table \ref{psr}) to search for their optical counterparts.  
For none of these pulsars has a proper motion been measured, so that the actual uncertainties on the pulsar positions at the epochs of our optical observations are somewhat larger than expected
 from the formal errors alone. 
 of our observation, and $v_{100}$, $d_{100}$ are the pulsar transverse velocity and distance in units of 100 km s$^{-1}$ and 100 pc, respectively.
However, with the only exception of PSR\, J2043+2740 (see Testa et al.\ 2018), for all our targets the angular displacement due to the unknown proper motion would be, for the nominal pulsar 
distances and an average transverse velocity of 400 km s$^{-1}$ (Hobbs et al.\ 2004), between $\sim$0\farcs1 and $\sim$0\farcs2, i.e. smaller than the formal position error.
Nonetheless, since the actual transverse velocity is unknown and the pulsar distances are uncertain we allowed for a reasonable tolerance in searching for the pulsar candidate counterparts.
Sections of the pulsar fields are shown in Fig.\ref{fc}.

\subsection{PSR\, J1846+0919}

\begin{figure*}
\centering
\begin{tabular}{cc}
\subfloat[J1846+0919]{\includegraphics[width=8.1cm]{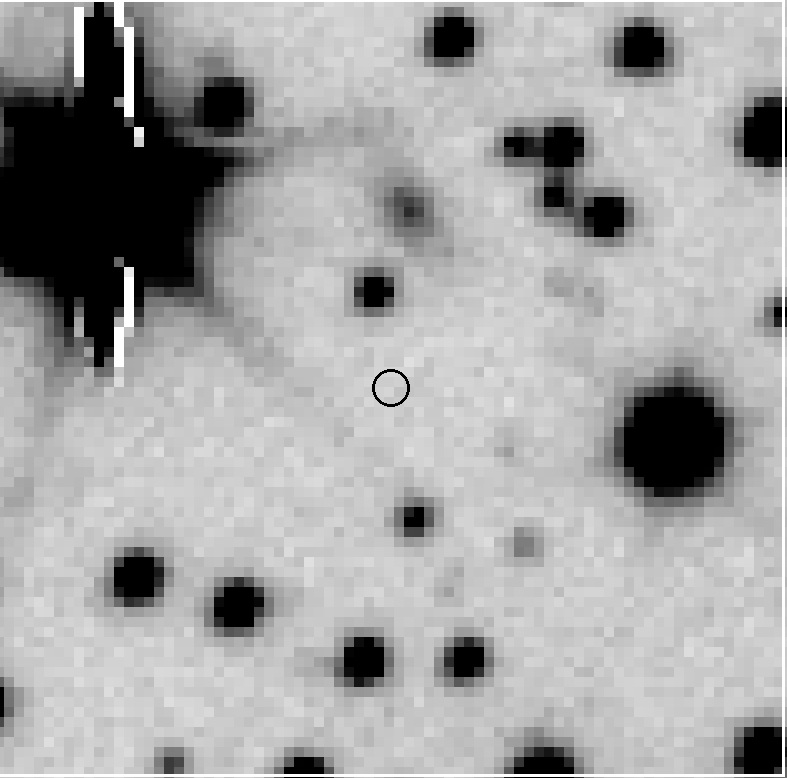}}  & 
\subfloat[J1907+0602]{\includegraphics[width=8cm]{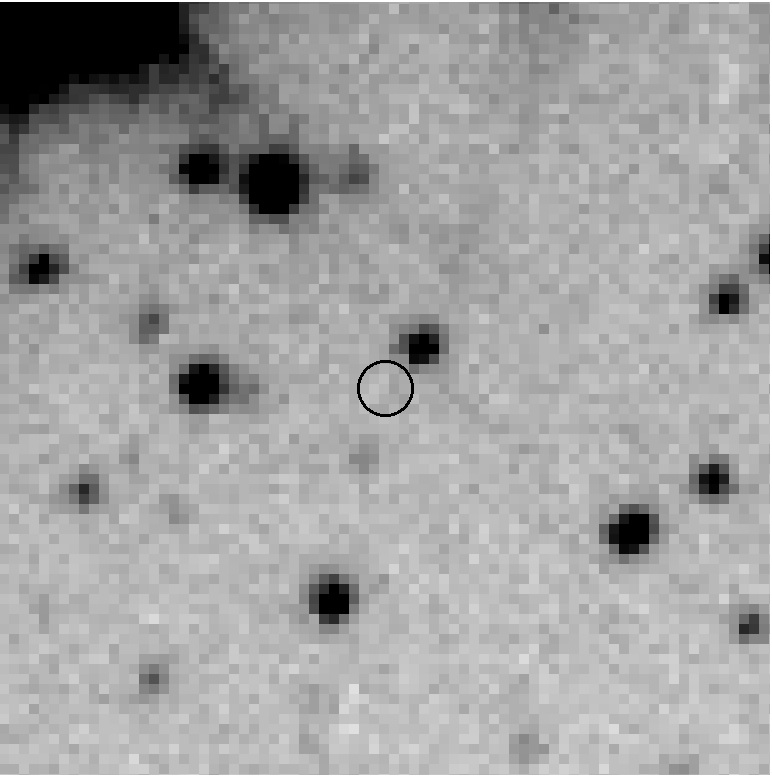}} \\  
\subfloat[J2043+2740]{\includegraphics[width=8cm]{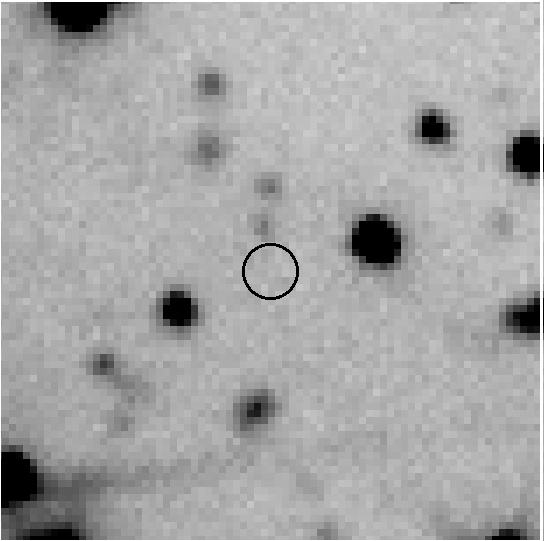}} & 
\subfloat[J2055+2539]{\includegraphics[width=8cm]{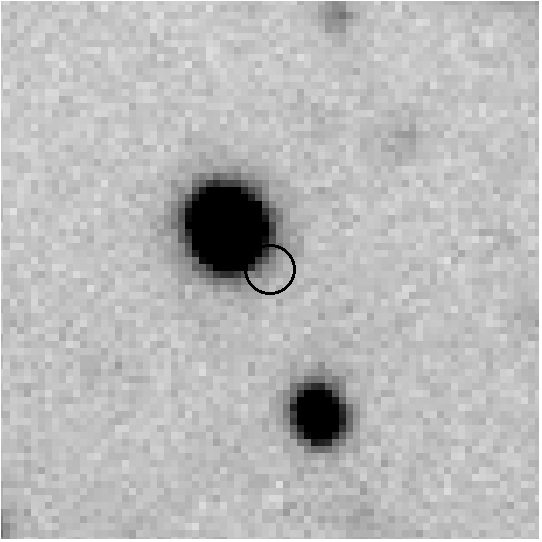}} \\  
\end{tabular}
\caption{\label{fc} 
$10\arcsec \times10\arcsec$ $r'$-band image sections around the fields of the four pulsars. The black circle corresponds to the $1\sigma$ error on the pulsar coordinates at the reference epoch (see Table \ref{psr}) and its size does not account for the $\la 0\farcs2$ systematic error on our absolute astrometry calibration. Apart from PSR\, J2043+2740, for all pulsars the estimated position uncertainty due to the unknown proper motion is well below the formal position error.}
\end{figure*}

No candidate counterpart is detected at the position of PSR\, J1846+0919 in the $g'$ and $r'$-band images, with the closest object ($g'=24.00\pm0.04$;  $r'=21.89\pm0.06$) 
being $\sim$ 2\farcs5 away, well beyond the $3\sigma$ uncertainty on the nominal pulsar position (Fig. \ref{fc}a).  Therefore, PSR\, J1846+0919 is still undetected in the optical. 
We computed the $3\sigma$ detection limits of our images from the rms of the sky background  at the pulsar position using the standard formula from Newberry (1991) with an aperture of radius equal to the image PSF and applying the aperture correction for consistency with the {\sc DAOPHOT-II}  photometry (Sectn. 2.2). We will adopt this convention throughout the rest of this paper.
The detection limits computed in this way are  $g'\sim 27.04$ and $r'\sim 26.04$
and we assumed these values as the upper 
limits on the PSR\, J1846+0919 optical brightness. These 
are the deepest constraints ever on the optical emission of this pulsar. 
Owing to the low count statistics of the \chan\ detection it was not possible to fit simultaneously for the spectrum and the  $N_{\rm H} $ (Marelli et al.\ 2015). 
However, the "pseudo distance" D$_{\gamma}\approx$1.4 kpc  would imply an $N_{\rm H} \approx 2 \times 10^{21}$cm$^{-2}$ after scaling the Galactic value in the pulsar
 direction (Marelli et al.\ 2015),  which is consistent with the estimates based on the correlation of He et al.\ (2013). From the Predehl \& Schmitt (1995) relation 
this would correspond to a reddening $E(B-V)\approx 0.33$.

\subsection{PSR\, J1907+0602 }

\begin{figure}
\centering
\begin{tabular}{cmd}
{\includegraphics[width=8.1cm]{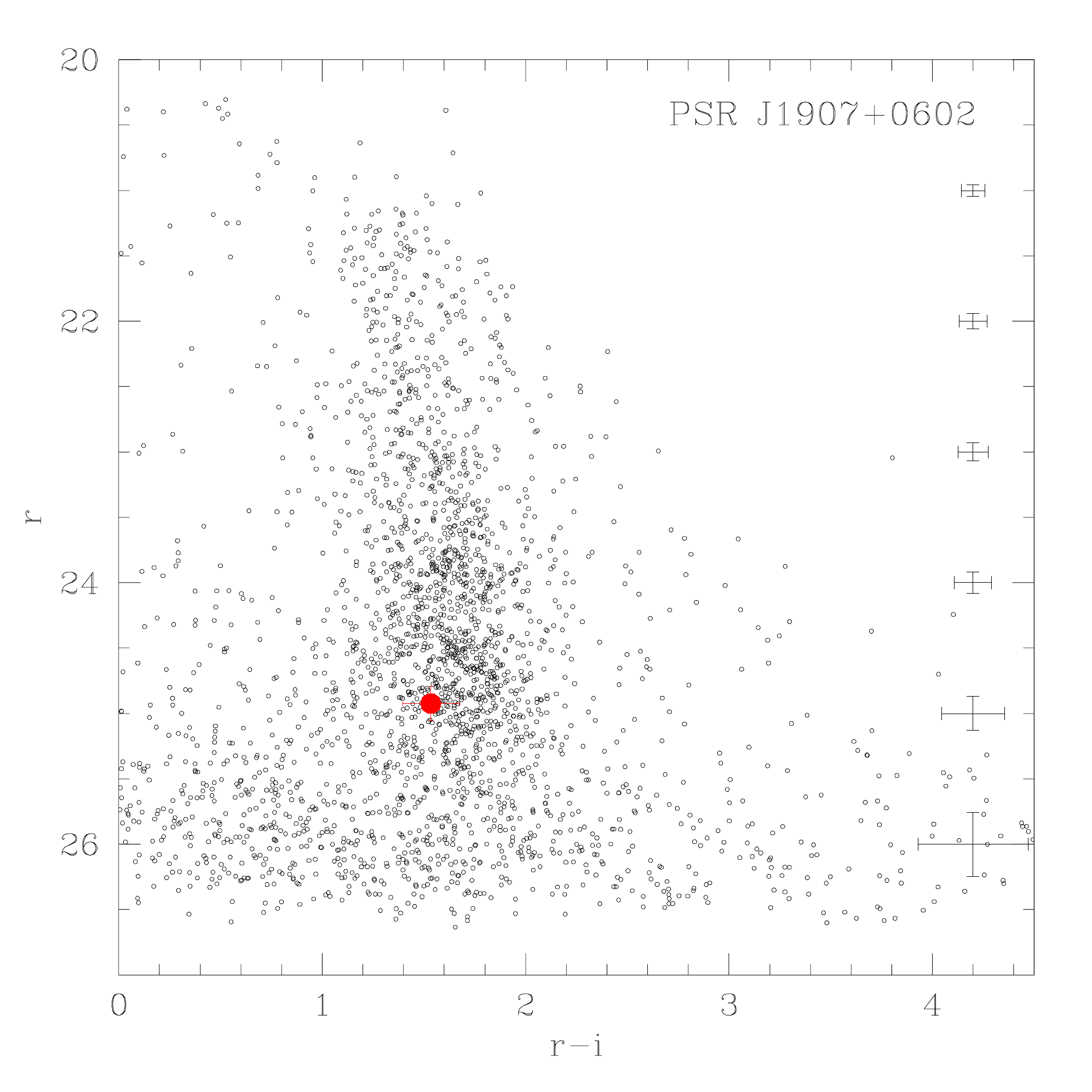}}  
\end{tabular}
\caption{\label{cmd1907} 
Observed $r'$, $r'-i'$ colour-magnitude diagram for all the stars detected in the PSR\, J1907+0602 field (black dots). The location of the object detected closest to the pulsar position (Fig.\ 1b) is marked by the large red dot. The points on the right of the diagram represent the average errors per magnitude bin.}
\end{figure}

The GTC image of the PSR\, J1907+0602 field is much more populated with stars than the VLT ones of Mignani et al.\ (2016c), probably owing to the fact that our $r '$ and $i'$-band 
observations mitigate the effects of the Galactic extinction in the pulsar direction. Indeed, while no object was found within a $10\arcsec \times 10\arcsec$ area around the pulsar 
position in their B and V-band images (see Fig.\ 1 of Mignani et al.\ 2016c), we found several objects in the same area (Fig. \ref{fc}b). In particular, one of them is close to the 
northern edge of the \chan\ error circle, 
at $\sim 1\farcs4$ from the pulsar position, detected both in the $r'$ and $i'$ bands. The object magnitudes are  $r'=24.93\pm0.13$ and  $i'=23.39\pm0.05$.  
We cannot claim an association with the pulsar based on the loose positional coincidence, 
for which we estimate a chance coincidence probability of $\sim 0.21$.
This is estimated as $1-\exp(-\pi\rho r^2)$, where   $r=1\farcs4$ is the angular distance between the object and the nominal pulsar position and $\rho$ is the number density of objects measured on the image. Nonetheless, it is still worth investigating whether 
this object can be considered a possible counterpart to the pulsar. Taking as a reference its $r'$-band magnitude 
would imply an optical luminosity $L_{\rm opt} \sim (2.6\pm0.7) \times 10^{30} d_{2.58}^2$ erg s$^{-1}$, where $d_{2.58}$ is the pulsar distance scaled to the DM-based value 
of 2.58 kpc (Yao et al.\ 2017). As in Mignani et al.\ (2016c), we estimated the interstellar reddening to the pulsar to be $E(B-V) = 0.88^{+0.12}_{-0.09}$ according to  
the $N_{\rm H}=41.1^{+3.5}_{-3.0} \times 10^{20}$ cm$^{-2}$ obtained from the fits to the X-ray spectrum (Abdo et al.\ 2013). We computed the interstellar extinction correction
in the $r'$ band from the coefficients of Fitzpatrick (1999). The optical luminosity would imply an emission efficiency $L_{\rm opt}/\dot{E} \sim 9.2 \times 10^{-7}$. This value is in 
the range observed for some of the young (kyr-old) pulsars identified in the optical (e.g., Moran et al.\ 2013) and is compatible with the constraints obtained for pulsars of age comparable to  PSR\, J1907+0602 (Paper I), although it is a factor of $\sim$500 higher than the efficiency of the 11 kyr-old Vela pulsar. However, this would not be unrealistic, in principle. Vela is still the only pulsar in the few ten kyr age range that  has been identified in the optical and using it as a yardstick to predict the optical efficiency of pulsars in the same age range, like  PSR\, J1907+0602, calls for the due caution. Therefore, we cannot rule out a priori that the object nearest to PSR\, J1907+0602 is a potential counterpart.  
Its non-detection  in the VLT images of Mignani et al.\ (2016c) implies $r'-V \la  -2$ and $r'-B \la -2.7$, after accounting for the difference between the AB and Vega systems, confirming that its colour ($r'-i' = 1.54\pm0.14$) indicates an intrinsically 
red spectrum  that cannot be the result of the differential extinction between the $r'$ and $i'$ bands  ($A_{r}-A_{i} \sim 0.66$). Such a spectrum would not be compatible with the 
flattish slope of the power-law (PL) optical spectra $F_{\nu} \propto \nu^{-\alpha}$ of the Crab and Vela pulsars (e.g., Mignani et al.\ 2007).  Indeed, as tentative as it may be 
based on two flux measurements only, a fit with a PL gives $\alpha \approx 4$, far too steep with respect to the values observed for other young or middle-aged pulsars for
which $\alpha\sim$0--1 (Mignani 2011). In order to further investigate the object's characteristics we compared its $r'-i'$ colour with that of stars in the GTC image.  
We ran the object detection in the $r'$ and $i'$-band images with {\sc DAOPHOT-II} and computed the photometry following the procedure described in Sectn.\ 2.2.
We considered all detections within the field of view of the  OSIRIS Chip 2 down to a 3 $\sigma$ limit above the sky background. 
We then matched the object catalogues obtained for each band to build the  colour--magnitude diagram (CMD). 
The observed, i.e. not corrected for the extinction, $r'$, $r'-i'$ CMD for all the stars detected in the PSR\, J1907+0602 field is shown in Fig. \ref{cmd1907}. 
As can be seen, the object location in the diagram is close to the main sequence and  has no peculiar characteristic that can support its association with PSR\, J1907+0602.
Therefore,  we conclude that it is, almost certainly, a field star.
No other object is detected close to the pulsar position down to $3\sigma$ limiting magnitudes of $r' \sim26.15$  and $i' \sim26.30$.

Like in Mignani et al.\ (2016c), we found no evidence of extended structures  around the pulsar, which can be associated with the PWN detected at TeV energies. Given its large angular extent ($\sim 40\arcmin$), however, only a very small part of the PWN  ($\approx4\%$) is covered by the combined field of view of the two chips of the OSIRIS detector ($7\farcm8 \times 7\farcm8$). For this reason, any attempt to determine a significant upper limit on the optical surface brightness of the PWN would be biased by the incomplete sampling of the sky background in the PWN region.

\begin{figure*}
\centering
\begin{tabular}{cmd}
{\includegraphics[width=8.1cm]{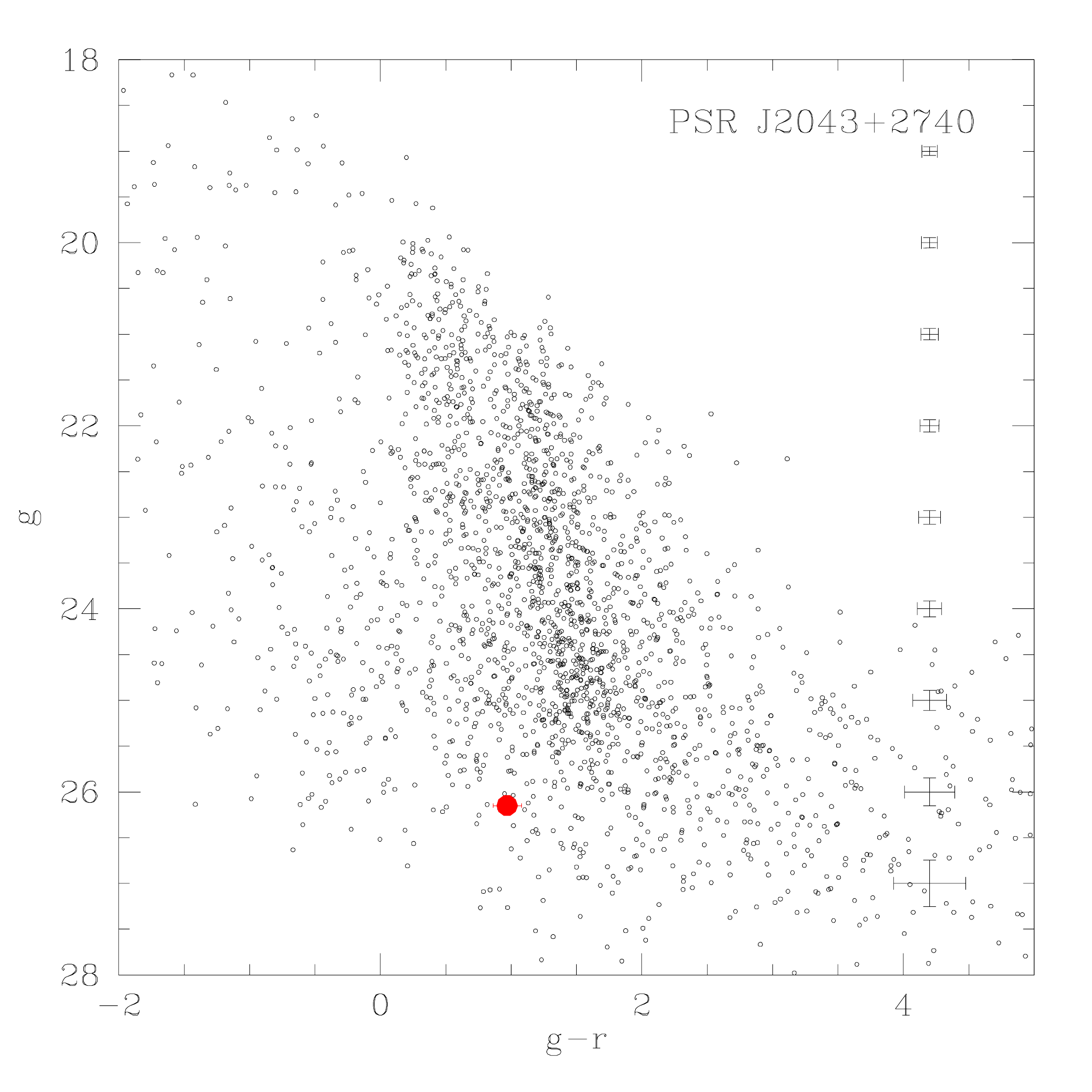}}  
{\includegraphics[width=8.1cm]{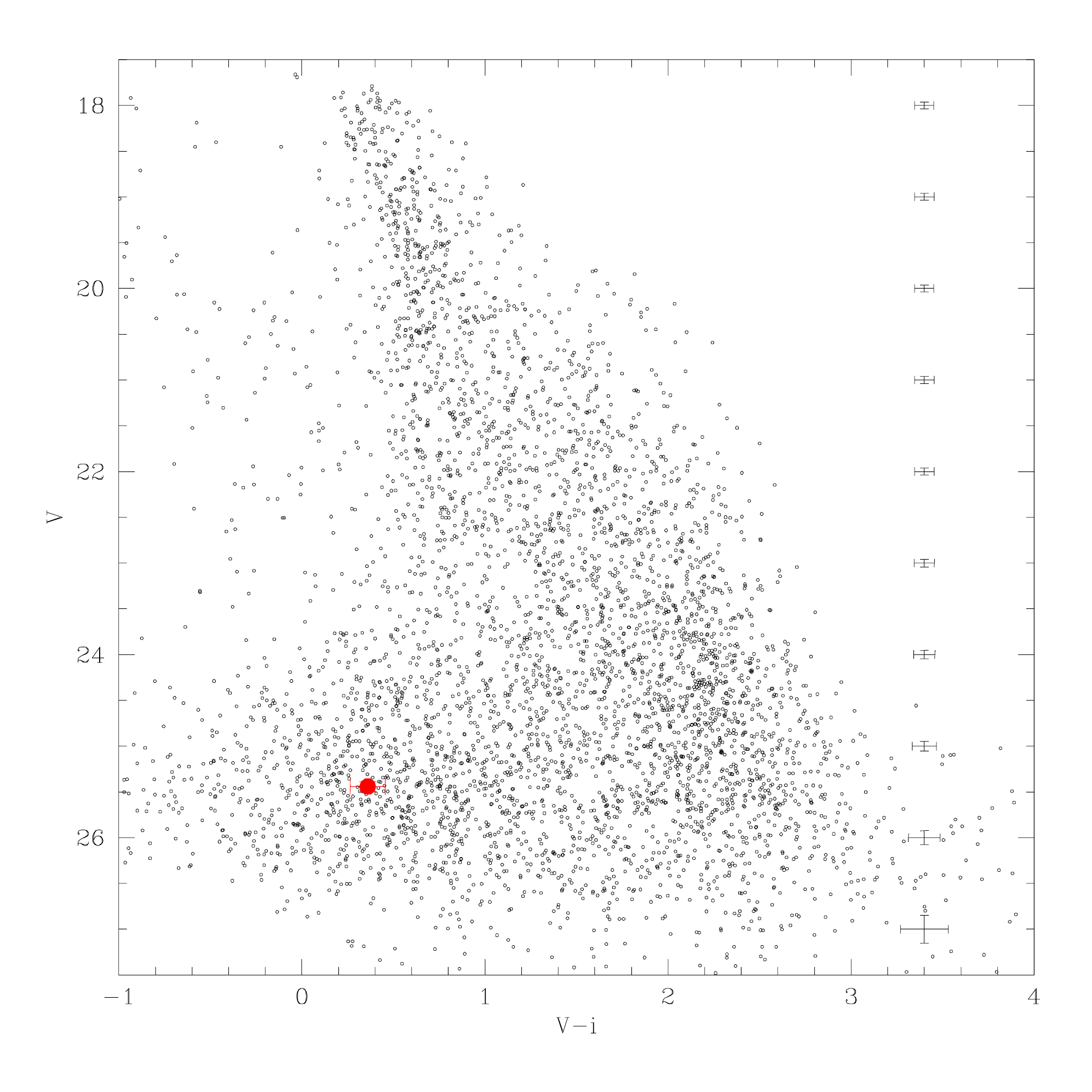}}  
\end{tabular}
\caption{\label{cmd2043} 
Left: observed $g'$, $g'-r'$ colour-magnitude diagram for all the stars detected in the PSR\, J2043+2740 field (black dots) obtained from the GTC/OSIRIS observations. 
Right: observed $V$, $V-i$ CMD obtained from the LBT observations of Testa et al.\ (2018) with the Large Binocular Camera (LBC). The location of the object detected closest to the pulsar position (Fig.\ 1c) is marked by 
the large red dot. The difference in the number of stars in the two diagrams is due to the different field of view of the two detectors, with the LBC one being almost 20 times as large. Like in Fig.\ 2, the points on the right of the diagrams represent the average errors per magnitude bin.  }
\end{figure*}

\subsection{PSR\, J2043+2740}

The object detected closest to the PSR\, J2043+2740 position (Fig. \ref{fc}c)  is that investigated in Testa et al.\ (2018) using LBT observations, and considered an unlikely pulsar candidate counterpart based on a colour analysis. Its $g'$ and $r'$-band magnitudes are  $26.35 \pm 0.10$ and $25.17 \pm 0.08$, respectively, corresponding to $g'-r' = 1.18\pm0.13$. This confirms that it has a quite red spectrum\footnote{The object has a $V-i = 0.36\pm0.09$, as measured in the LBT images of Testa et al.\ (2018).}, contrary to what one would expect for a neutron star. The observed  $g'$, $g'-r'$  CMD of the field stars is shown in Fig. \ref{cmd2043}a. The diagram has been built using the same procedure applied in Sectn.\ 3.2. The object of Testa et al.\ (2018) lies at the faint end of the main sequence and its colour is redder than the bulk of the sequence but comparable to other stars in the field of view. For comparison, we also present the unpublished $V$, $V-i$ CMD obtained from the LBT observations (Fig. \ref{cmd2043}b), which shows that the object location with respect to the main sequence is consistent in both diagrams.  This confirms the conclusion that it is a field star. No new candidate counterpart has been detected in the GTC $g'$ and $r'$-band images of PSR\, J2043+2740 (Fig. \ref{fc}c).  The $3\sigma$ limiting magnitudes are $g'\sim27.26$ and $r'\sim26.69$. To force the detection limit, we co-added the  $g'$ and $r'$-band GTC images, as well as the  $U, V, i$-band LBT ones, but in both cases we found no evidence of a new detection at the pulsar position.  As in Testa et al.\ (2018), in the following discussion we assume an $E(B-V)\la 0.06$, derived from $N_{\rm H} \la 3.6 \times 10^{20}$ cm$^{-2}$ (Abdo et al.\ 2013), as a reference value for the interstellar reddening.

\subsection{PSR\, J2055+2539}

Finally, for PSR\, J2055+2539 the pulsar position falls in the PSF wings of a relatively bright star at $\sim 2\arcsec$  (Fig. \ref{fc}d). 
This is the same star observed by Beronya et al.\ (2015)  in snapshot observations 
with the Bolshoi Teleskop Alt-azimutalnyi (BTA) 6m telescope (Special Astrophysical Observatory, Russia).
However, the star magnitude ($r'=20.34\pm0.04$) is far brighter than expected 
for a middle-aged isolated neutron star at a distance of 0.6 kpc and with an interstellar reddening $E(B-V)\sim0.4$, as obtained from 
an $N_{\rm H} = (2.18\pm0.26)\times 10^{21}$cm$^{-2}$ (Marelli et al.\ 2016).  This would imply an exceptionally large emission efficiency, 
$L_{\rm opt}/\dot{E} \sim 10^{-4} d_{0.6}^{2}$,  where $L_{\rm opt}$ refers to the $r'$ band and $d_{0.6}$ is the pulsar distance in units of 0.6 kpc,
at least two orders of magnitude larger than that of other middle-aged pulsars detected in the optical (see, e.g. Moran et al.\ 2013).  
Regardless of emission efficiency arguments, the angular separation between the star and the pulsar ($\sim2\arcsec$), 
too large to be accounted for by its unknown proper motion, argues against an association. Indeed, since the epoch of 
the reference \chan\ position (MJD=57277) is very close in time to that of our GTC observations (MJD= 57572) the effect of the
unknown pulsar proper motion on its expected position would be only $\sim$0\farcs1, for the assumed pulsar distance (0.6 kpc) 
and transverse velocity (400 km s$^{-1}$). This means that the pulsar should move much faster and/or be much closer to account 
for the yearly angular displacement ($\sim 2\farcs5$ yr$^{-1}$) required to make a positional association  possible.  However, such a large
displacement  is ruled out by the comparison between the star coordinates measured on the GTC and  DSS-2 images of the field (epoch 1995.55). 
Accounting for their associated absolute astrometry errors, this implies an angular displacement $\le 0\farcs3$ ($1 \sigma$) over a time span of $\sim$21 years, 
corresponding to a proper motion of $\le$ 15 mas yr$^{-1}$ for this star. Therefore, we can certainly rule it out  as a candidate counterpart 
to PSR\, J2055+2539. The pulsar is, then, undetected down to $g'\sim26.83$ and $r'\sim26.46$, 
where the limiting magnitudes are affected by the presence of the relatively bright nearby star. Nonetheless, ours are the deepest
limits on the pulsar flux obtained so far, much deeper than those  ($V\sim22.78$; $R\sim23.22$) of Beronya et al.\ (2015).

\begin{table}
\centering
\caption{Computed detection limits at the pulsar positions. The table also reports the assumed distance (see Section 1) and reddening along the line of sight estimated from the $N_{\rm H}$ (Predehl\& Schmitt 1995) obtained from the X-ray spectral fits (Abdo et al.\ 2013; Marelli et al.\ 2016).  For  PSR\, J1846+0919, the $N_{\rm H}$ has been computed by scaling the Galactic value for the pulsar "pseudo-distance" D$_{\gamma}$ (Marelli et al.\ 2015).}
\label{ul}
\begin{tabular}{lllccl} \hline
 Name      &  \multicolumn{2}{c}{Limits ($3\sigma$)} &     d       & $N_{\rm H}$ &E(B-V) \\ 
                 &                              $g'$ & $r'$                                  & (kpc)  & ($10^{20}$ cm$^{-2}$)         & \\ \hline
	J1846+0919   & $27.04$ & $26.04$ & 1.4 & 20 &0.33 \\
	J1907+0602   &  $26.30$$^{\dagger}$  & $26.15$  & 2.58 & $41.1^{+3.5}_{-3.0}$&$0.88^{+0.12}_{-0.09}$ \\
	J2043+2740   & $27.26$ &  $26.69$ &  1.48 & 3.6  &$0.06$\\
	J2055+2539   & $26.83$ & $26.46$   & 0 .6 &$21.8\pm 2.6$ &0.4 \\ \hline
\end{tabular}
\vspace{0.5cm}
$^{\dagger}$ This limit is  in the $i'$ band (see Section 3.2).
\end{table}

\section{Discussion}

Table 3 summarises the detection limits for the four pulsars discussed in this work along with the assumed values of distance, $N_{\rm H}$, and inferred  interstellar reddening. We remind that for the radio-silent PSR\, J1846+0919 and PSR\, J2055+2539 we assumed the "pseudo-distance"  D$_{\gamma}$ (Marelli et al.\ 2015; 2016), whereas for  the radio-loud pulsars PSR\, J1907+0602  and PSR\, J2043+2740 we assumed the DM-distance based on the Yao et al.\ (2017) model instead of that of Cordes \& Lazio (2002). This explains the  difference between the distance values in Table 3 and those reported in the 2PC (Abdo et al.\ 2013).

All  the four pulsars remain undetected at optical wavelengths down to $3 \sigma$ limits of $g' \approx 27$.  These are close to the magnitudes expected for 
the three middle-aged pulsars PSR\, J1846+0919, PSR\, J2055+2539, PSR\, J2043+2740  by scaling for the $\dot{E}$, distance, and reddening the observed optical brightness of pulsars of 
similar characteristic age. This might also be true 
for   PSR\, J1907+0602, for which this kind of extrapolations, based on the Vela pulsar case only, are way more uncertain (see Section 3.2). Therefore, we cannot rule out that our targets  
might be eventually detected in deeper observations carried out with the same telescope/instrument set-up exploiting, e. g. a factor of two improvement in the integration time and seeing conditions. 

Their non-detection in the current observations may reflect the uncertainty in assuming the brightness of other pulsars as a reference, with only fifteen or so identified in the optical  over all age ranges (Mignani 2011), 
but may also imply larger distance and reddening values than inferred from current radio, $\gamma$ and X-ray data. 
As a matter of fact, both DM-based distances for radio-loud pulsars (Yao et al.\ 2017) and $\gamma$-ray "pseudo distances" 
(Saz Parkinson et al.\ 2010) for the radio-quiet ones can be very uncertain, with the uncertainty difficult to quantify a priori, especially in the latter case. 
This calls for more precise distance indicators. As of now, only for a tiny fraction of  the radio-loud $\gamma$-ray pulsars have parallax measurements
been obtained (see, e.g. Abdo et al.\ 2013), whereas for the radio-quiet ones a parallax measurement (not yet feasible in the X-rays, let alone in $\gamma$-rays) 
can only follow the optical identification. This means that the lack of a direct distance measurement  makes the optical identification of radio-quiet $\gamma$-ray pulsars 
somewhat more difficult, although not hopeless, as shown by the case of Geminga (Bignami et al.\ 1987) and of other classes of radio-quiet INSs (e.g. Mignani 2011).
Concerning the reddening, most estimates rely on the $N_{\rm H}$ inferred from the fits to the X-ray spectra, which are to some extent model-dependent, or extrapolated 
from the Galactic value  for the (in turn uncertain) pulsar distance when no, or little spectral information is available (e.g. for PSR\, J1846+0919). Indeed, about half of the $\gamma$-ray pulsars in the 2PC  (Abdo et al.\ 2013) have 
either no X-ray identification or a poorly constrained  X-ray spectrum. 

Together with more precise and bias-free distance and reddening estimates, proper motion-corrected coordinates are equally important to minimise the uncertainty on the pulsar position and the number of spurious matches, a clear limitation for optical follow-ups, especially in crowded fields. The case of PSR\, J2043+2740 is an example of the importance of an updated position, either from radio or \chan\ observations, with its reference coordinates dating back to 1995 (Ray et al.\ 1996), resulting in an uncertainty of a few arcsec on the pulsar position at the epoch of our observations (see, also Testa et al.\ 2008).

\begin{figure*}
\centering
\begin{tabular}{cc}
\subfloat[J1846+0919]{\includegraphics[width=8.9cm]{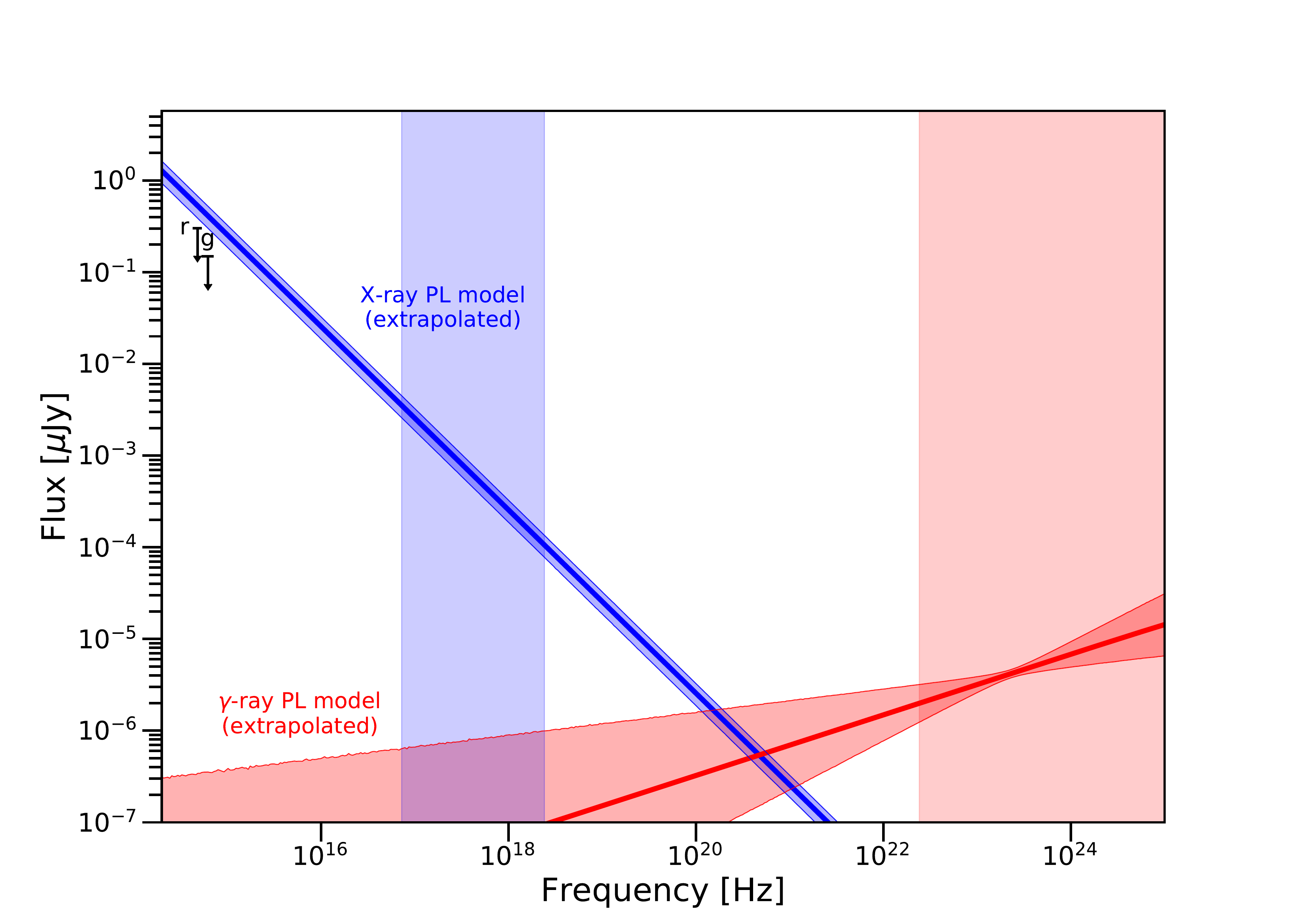}}   &
\subfloat[J1907+0602]{\includegraphics[width=8.9cm]{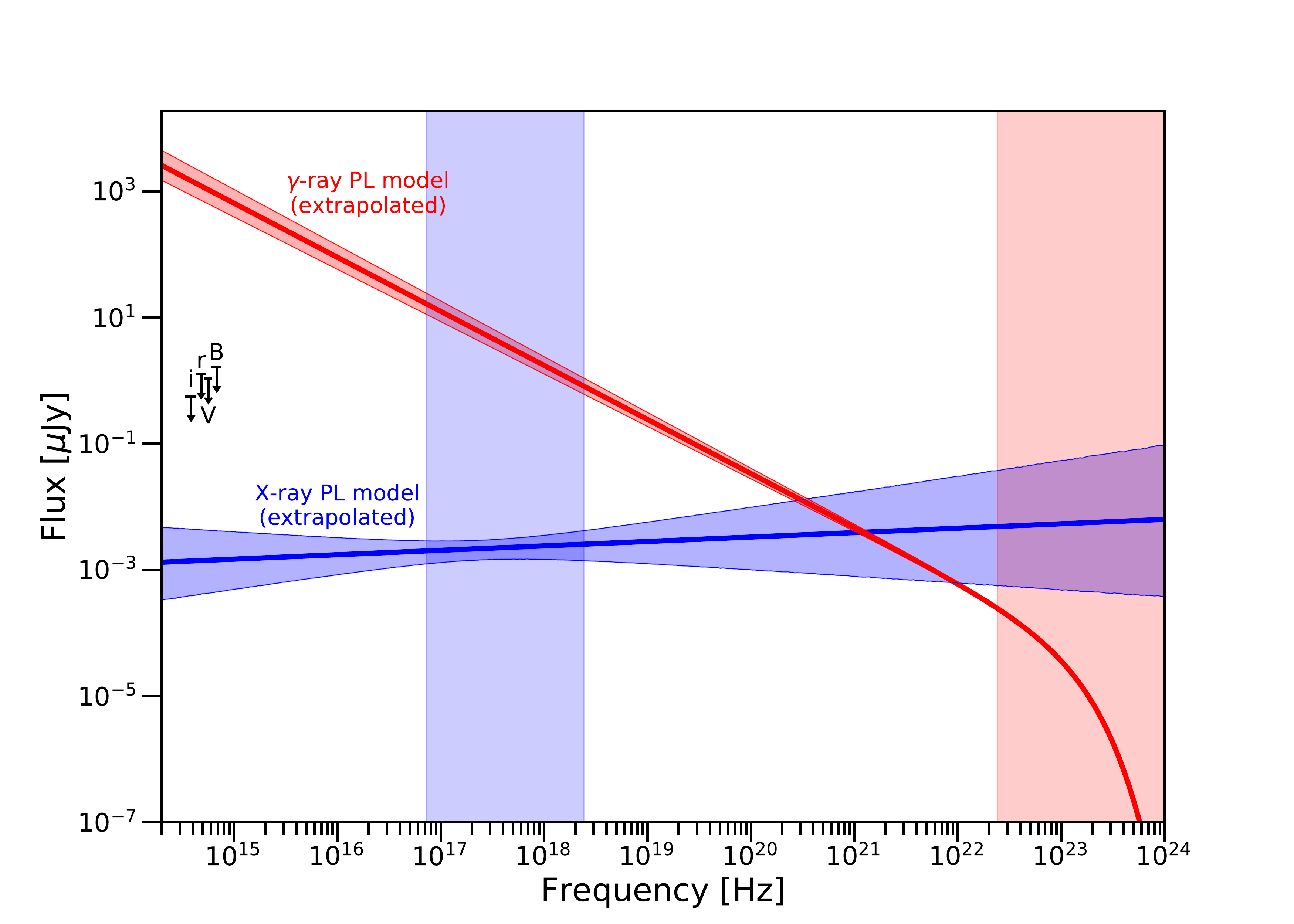}}  \\
\subfloat[J2043+2740]{\includegraphics[width=8.9cm]{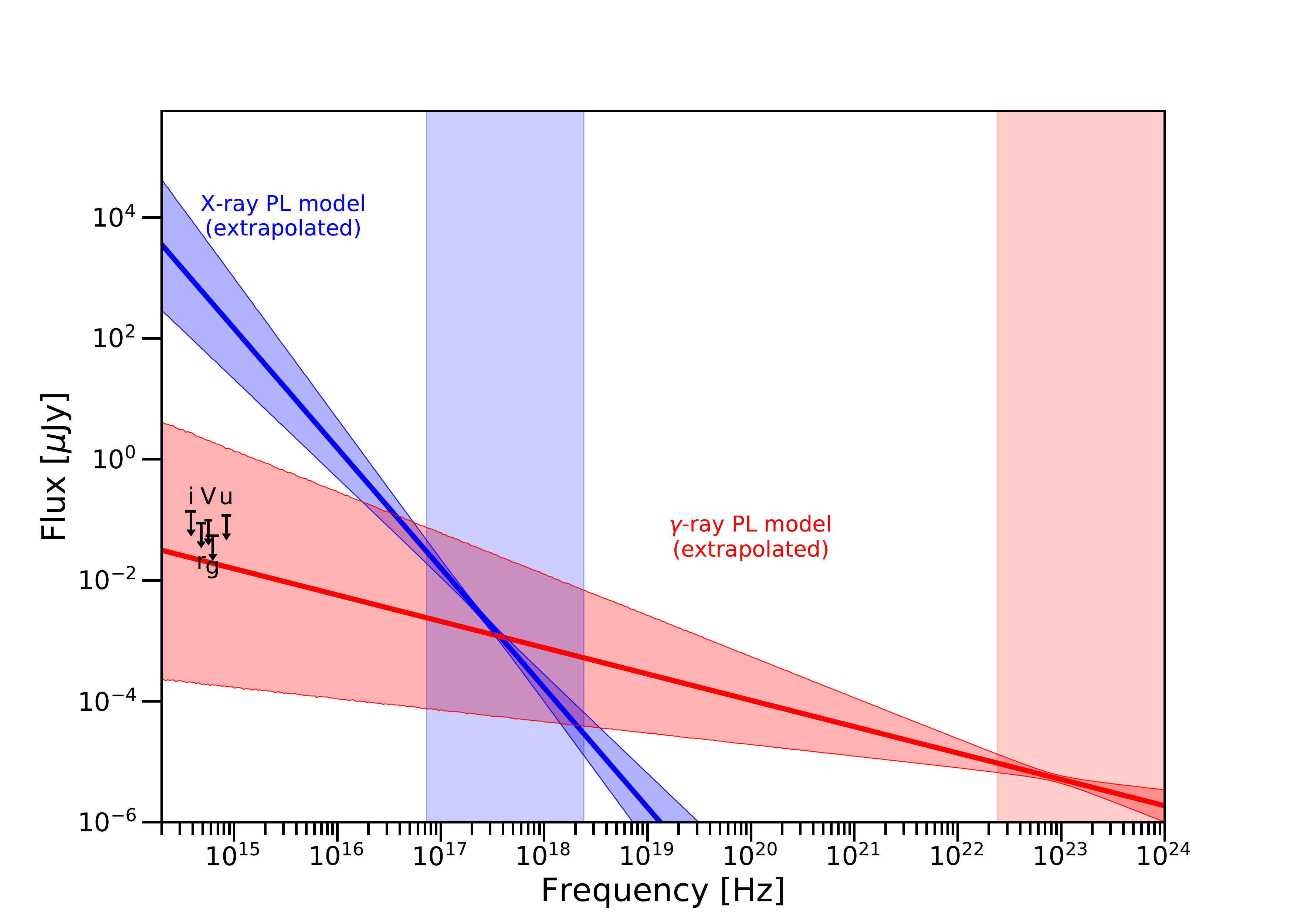}}  &
\subfloat[J2055+2539]{\includegraphics[width=8.9cm]{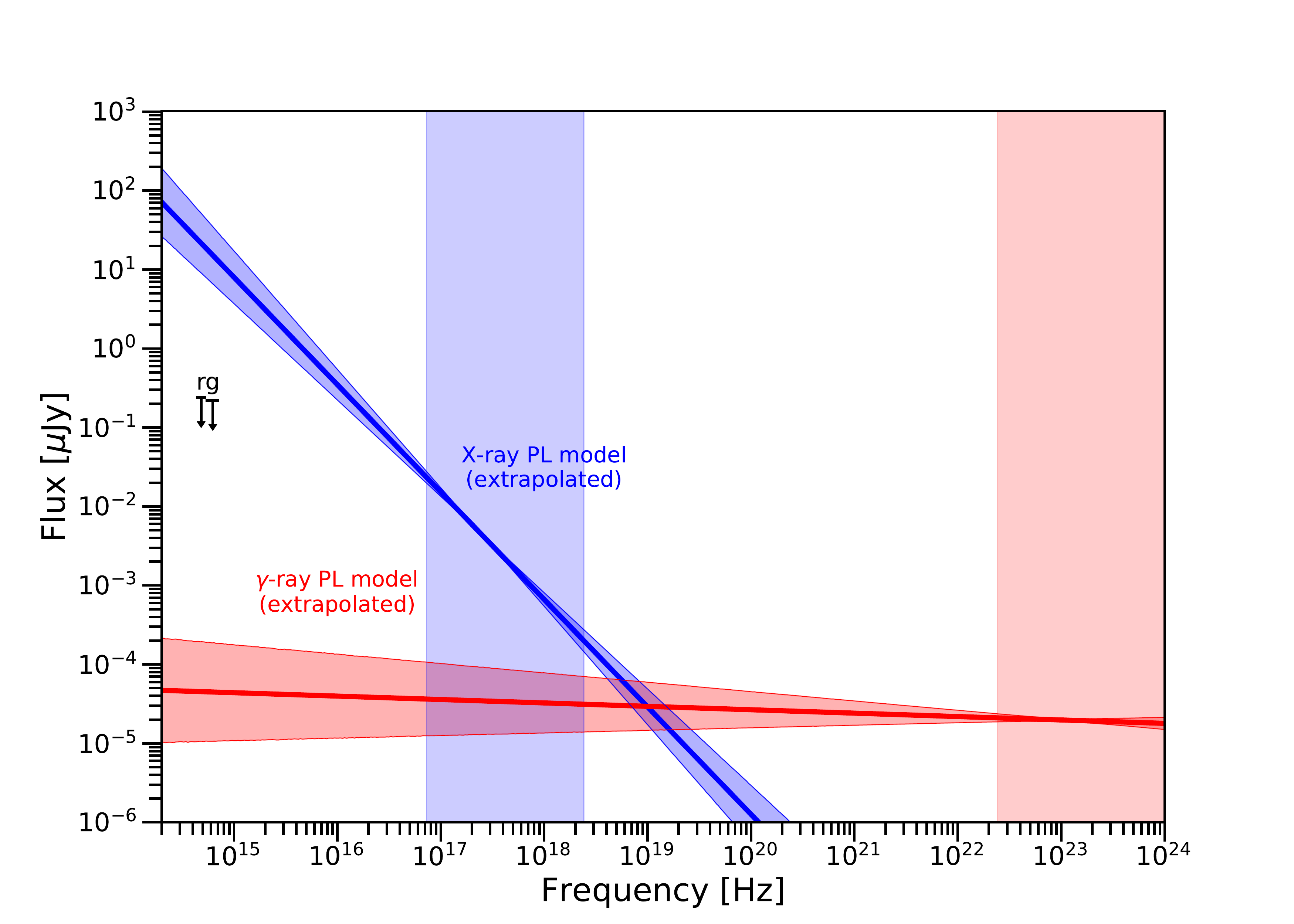}}  \\
\end{tabular}
\caption{\label{sed} 
Multi-wavelength SEDs of the four pulsars discussed in this work. The extrapolations of the $\gamma$ and X-ray PL spectra are drawn in the solid red and blue lines, respectively. The red and blue shaded areas around these lines correspond to the $1 \sigma$ uncertainty on the extrapolation. The $\gamma$-ray spectral parameters are taken from the 3FGL catalogue (Acero et al.\ 2015) and the X-ray ones from the 2PC (Abdo et al.\ 2013). For PSR\, J2055+2539, however, we assumed the most recent X-ray spectral parameters of Marelli et al.\ (2016), whereas for PSR\, J1846+0919 we fixed the X-ray photon index $\Gamma_{\rm X}$ to 2 (Marelli et al.\ 2015). The blue and red vertical bands mark the spectral regions were the X and $\gamma$-ray spectra have been measured. The computed optical flux upper limits are indicated by the arrows and labelled by the filter names. For PSR\, J1907+0602 and PSR\, J2043+2730  the $B,V$ and $U,V, i$-band upper limits from Mignani et al.\ (2016c) and Testa et al.\ (2018), respectively, are also plotted. }
\end{figure*}

From our detection limits (Table \ref{ul})  we computed the corresponding spectral flux upper limits corrected for the interstellar extinction using the most conservative reddening value and the coefficients of Fitzpatrick (1999).  We then compared our flux upper limits with those existing in the literature and with the extrapolations of the  X and $\gamma$-ray 
PL spectra  to search for low-energy turnovers in the SED of the radiation emitted from the pulsar magnetosphere. For both PSR\, J1846+0919 and PSR\, J2055+2539, our results make this investigation possible 
for the first time.
We assumed the $\gamma$-ray spectral parameters (photon index $\Gamma_{\gamma}$ and cut-off energy $E_{\rm c}$) and  100 MeV--100 GeV $\gamma$-ray fluxes from the {\em Fermi}  LAT Third Source Catalogue (3FGL; Acero et al.\ 2015), while we assumed the X-ray spectral parameters (photon index $\Gamma_{\rm X}$)  and unabsorbed  0.3--10 keV X-ray fluxes from the Second {\em Fermi} LAT Catalogue of $\gamma$-ray pulsars  (2PC; Abdo et al.\ 2013),  with the exception of PSR\, J1846+0919 and PSR\, J2055+2539 for which we assumed those published in Marelli et al.\ (2015) and Marelli et al.\ (2016), respectively. 
For both J2043+2740  and PSR\, J2055+2539, the addition of a blackbody (BB) emission component to the PL, associated with thermal radiation from the neutron star surface, does not improve significantly  the fit quality (Testa et al.\ 2018; Marelli et al.\ 2016). Therefore,  we assumed a simple PL spectrum. 
The optical spectrum of these four pulsars is obviously unknown. For the young PSR\, J1907+0602, the optical emission is expected to be non-thermal and described by a PL, like in the case of the Vela pulsar (e.g., Mignani 2011). For the middle-aged pulsars, the optical emission is ascribed to both non-thermal and thermal emission processes but we cannot disentangle the contribution of the two components, a PL and a Rayleigh-Jeans (RJ), from our upper limits. However, according to the known spectra of middle-aged pulsars, the RJ component seems to become dominant towards the UV (e.g. Mignani 2011), so that we can assume that the optical emission of middle-aged pulsars in the $\sim 4000$--8000 \AA\  wavelength range is mostly non-thermal.

As can be seen, for the three middle-aged pulsars the optical flux upper limits are below the extrapolation of the X-ray PL spectrum, indicating a spectral break at low energies.  These breaks are observed in some of the other middle-aged pulsars with an identified optical counterpart, i.e.  PSR\, B1055$-$52 (Mignani et al.\ 2010), PSR\, J1741$-$2154 (Mignani et al.\ 2016a), and Geminga, where the PL component to the optical spectrum is below the extrapolation of the X-ray PL one (Kargaltsev \& Pavlov 2007), although only marginally in the latter case. However, this is not the case for PSR\, B0656+14 (Kargaltsev \& Pavlov 2007), where the extrapolation of the X-ray PL component match, within the error, the optical one. For another middle-aged $\gamma$-ray pulsar, PSR\, J0357+3205, the GTC upper limits are also below the X-ray PL extrapolation (Kirichenko et al.\ 2014).  With the exception of PSR\, J2043+2740, the optical flux upper limits for both PSR\, J1846+0919 and PSR\, J2055+2539 are well above the extrapolation of the $\gamma$-ray PL,  which means that the optical spectrum might be compatible with the $\gamma$-ray one.  Only for PSR\, J2043+2740, however,  this hypothesis can be realistically verified with a factor of 100 more sensitive optical observations, perhaps still within reach of the next generation of 30--40m-class telescopes.  
For PSR\, J1907+0602 the new optical flux upper limits confirm the presence of a spectral break of the $\gamma$-ray PL  at low energies of Mignani et al.\ (2016c) and prove, once again, that the SEDs of pulsars in the few ten kyr age range do not follow a unique template, as already pointed out in Paper I.

To summarise, our results seem to suggest, with the due caution owing to the still limited sample, that the non-thermal SEDs of middle-aged $\gamma$-ray pulsars are more similar to each other than the SEDs of pulsars in the ten kyr age range, possibly pointing at an evolutionary effect.  Although tantalising, such comparisons come with due caveats. The first one is that extrapolating the $\gamma$ and X-ray PLs over ten and four orders of magnitudes in energy, respectively, might be an oversimplification and more advanced emission and spectral models might be necessary to fully account for the multi-wavelength spectra from the pulsar magnetosphere. For instance,  recent modelling of the pulsar multi-wavelength emission (Torres  2018)  shows that the X-ray PL does not follow the simple extrapolation of the $\gamma$-ray one (see e.g., Figure 2 of this paper). This conclusion is in line with the SEDs of the four pulsars discussed in this work (Figure \ref{sed}).
There might be some exceptions though, e.g. PSR\, J1048$-$5832 (Razzano et al.\ 2013), where the extrapolation of the $\gamma$-ray PL matches, within the errors, the X-ray one.  The second caveat is that,  in some cases, the contribution of  emission components other than that from the pulsar magnetosphere, e.g. thermal emission from the neutron star surface in the  X-rays or non-thermal emission from an unresolved PWN in the X and $\gamma$-rays,  cannot be easily accounted for. This makes the extrapolation of the pulsar high-energy PL spectra at optical wavelengths more uncertain.

Table \ref{lum} shows the computed upper limits on the optical luminosity $L_{\rm opt}$ (in the $r'$ band for a direct comparison between the four pulsars), optical emission efficiency
$\eta_{\rm opt} = L_{\rm opt}/\dot{E}$, and ratios between the unabsorbed $r'$-band  flux ($F_{\rm opt}$) and the X-ray ($F_{\rm X}$) and  $\gamma$-ray ($F_{\gamma}$) fluxes  in the 0.3--10 keV and 100 MeV--100 GeV energy ranges, respectively. We computed the optical luminosities in the $r'$ band from the unabsorbed  $r'$-band flux corrected for the interstellar reddening  (Fitzpatrick 1999) and scaled for the distance, assuming the values in Table \ref{ul}.
 Like we did for the SED analysis, we assumed  the pulsar 100 MeV--100 GeV $\gamma$-ray flux  from the 3FGL (Acero et al.\ 2015) 
and its unabsorbed non-thermal  0.3--10 keV X-ray flux from the 2PC (Abdo et al.\ 2013), while for PSR\, J1846+0919 and PSR\, J2055+2539 
we assumed the X-ray flux of Marelli et al.\ (2015) and Marelli et al.\ (2016), respectively.
Following the assumption on the pulsar optical spectrum in the 4000-8000 \AA\ wavelength range (see discussion earlier in this section),  we can assume that the pulsar flux in the $r'$ band  is mostly (if not entirely) non-thermal.

For both PSR\, J1907+0602 and PSR\, J2043+2740, the updated upper limits of the optical luminosity, efficiency and flux ratios are consistent, 
accounting for the difference between the GTC $r'$ filter and  other optical filters (the VLT v$_{\rm HIGH}$ and the LBT V-BESSEL), with the 
values published in Mignani et al.\ (2016c) and Testa et al.\ (2018). Therefore, previous conclusions are confirmed by our new observations. In particular, for PSR\, J1907+0602 we cannot rule out that this pulsar is a more powerful optical emitter than the Vela pulsar itself.
We note that for PSR\, J1907+0602 we have assumed the distance computed from the most recent Yao et al.\ (2017) model, whereas in Mignani et al.\ (2016c) we assumed that computed 
from the NE2001 model (Cordes \& Lazio 2002). This obviously influences the derived values of $L_{\rm opt}$ and $\eta_{\rm opt}$ and must be taken into account when these values are cross-compared.

For the other two pulsars, PSR\, J1846+0919 and PSR\, J2055+2539, (as well as for PSR\, J2043+2740), the upper limits on all the above quantities derived from our observations are 
consistent with the values measured for  middle-aged pulsars detected in the optical (e.g. Moran et al.\ 2013), suggesting that the pulsar multi-wavelength properties evolve smoothly above a certain value of the spin-down age.

\begin{table}
\centering
\caption{Computed  upper limits on the pulsar luminosity and efficiency in the $r'$ band for the distance and interstellar extinction values reported in Table  3 together with the upper limits on the unabsorbed optical--to--X-ray ($F_{\rm opt}/F_{X}$)  and optical--to--$\gamma$-ray ($F_{\rm opt}/F_{\gamma}$) flux ratios.  
}
\label{lum}
\begin{tabular}{lcccc } \hline
 Name      &  $L_{\rm opt}$ & $\eta_{\rm opt}$ &  $F_{\rm opt}/F_{X}$   & $F_{\rm opt}/F_{\gamma}$  \\ 
                 &   (erg s$^{-1}$) &                          &                                      &   \\ \hline
	J1846+0919  & 7.50$\times10^{28}$ & 		2.40$\times10^{-6}$	&  0.029 &1.38$\times10^{-5}$ \\		
		J1907+0602  & 1.11$\times10^{30}$ &		3.95$\times10^{-7}$	&  0.019 & 4.29$\times10^{-6}$  \\		
	J2043+2740  & 2.49$\times10^{28}$ &		4.45$\times10^{-7}$  &  0.004 &7.37$\times10^{-6}$ \\		
	J2055+2539  & 1.16$\times10^{28}$ & 		2.28$\times10^{-6}$	&  0.007 & 4.60$\times10^{-6}$ \\ \hline	
	\end{tabular}
\vspace{0.5cm}
\end{table}

\section{Summary and conclusions}

We have carried out deep optical observations of four $\gamma$-ray pulsars (Table 1) with the 10.4 m GTC, the first ever obtained for both PSR\, J1846+0919 and PSR\, J2055+2539.  We have examined possible candidates detected close to the reference pulsar positions but they all turned out to be field stars.  In all cases, we set deep limiting magnitudes ($g' \approx 27$), which we assume as upper limits on the pulsar brightness. 
The inferred limits on the optical luminosity, emission efficiency, multi-wavelength flux ratios of these four pulsars (Table 4) indicate that their multi-wavelength behaviour might be similar to that of other pulsars of similar characteristic age. The multi-wavelength SEDs of the three middle-aged pulsars features the same spectral breaks as observed in other middle-aged $\gamma$-ray pulsars identified in the optical, whereas the SED of PSR\, J1907+0602 confirms that pulsars in the few ten-kyr age range do not always feature  the same spectral breaks. Whether this points at an SED evolution with the pulsar characteristic age is yet unclear.
All in all, the comparison of the pulsar SEDs over different age ranges shows, once again, that deriving expectation values for the optical fluxes based on the extrapolation of the high-energy spectra is hazardous.

The detection limits for the four pulsars ($g' \approx 27$) are close to the magnitudes expected by scaling the optical brightness of pulsars of similar characteristic age. 
The non-detection of our targets reflects, to some extent, the uncertainty in assuming other pulsars as standard candles but also the still large uncertainties on  the pulsar distance and reddening, as we remarked in Sectn. 4,   In turn, this reflects a certain delay in performing deep follow-up observations in radio and X-rays after the pulsar detection in $\gamma$-rays. 
The situation is slowly improving (see, e.g. Marelli et al.\ 2015; Zyuzin et al.\ 2018), especially for the $\approx$100 more $\gamma$-ray pulsars detected since the publication of the 2PC (Abdo et al.\ 2013). This is due to the fact that the latest $\gamma$-ray pulsars  to be detected are, usually, the faintest ones and, presumably, also the faintest at other wavelengths, which limits the success of follow-up attempts. 

Comprehensive summaries of the X-ray and optical observations of the $\gamma$-ray pulsars detected by {\em Fermi} and of their astrophysical implications 
will be presented in companion papers to the Third {\em Fermi} LAT Catalogue of $\gamma$-ray pulsars (3PC; {\em Fermi}/LAT Collaboration 2018, in preparation).
According to the discovery rate of new $\gamma$-ray pulsars, with the 3PC presumably doubling the number of entries in the 2PC, follow-up observations are due to continue  in the next years, well after the end of the {\em Fermi}  mission. In the next decade, these observations will also exploit the advent of new facilities such as ESA's {\em Athena}, in the X-rays, ESO's ELT, in the optical, and the SKA in radio, which will push the detection limits from a factor of ten up to a few orders of magnitudes. In particular, SKA will make it possible to search for so far undetected radio pulsars among the increasing population of unidentified $\gamma$-ray sources  ($\sim 1000$ in the 3FGL; Acero et al.\ 2015) and  treasure the legacy of the {\em Fermi} mission.

\section*{Acknowledgments}
Based on observations made with the Gran Telescopio Canarias (GTC), installed in the Spanish Observatorio del Roque de los Muchachos of the Instituto de Astrofisica de Canarias,  in the island of La Palma. RPM acknowledges financial support from an INAF "Occhialini Fellowship". Researchers at ICE, CSIC have been supported by grants
AYA2015-71042-P, and SGR 2017-1383. NR is also supported by an NWO Vidi Grant. We thank the anonymous referee for his/her constructive comments to our manuscript.

\label{lastpage}

\end{document}